\documentstyle[twoside,fleqn,espcrc2]{article}

\newcommand{\be}{\begin{equation}}
\newcommand{\ee}{\end{equation}}
\newcommand{\bea}{\begin{eqnarray}}
\newcommand{\eea}{\end{eqnarray}}
\newcommand{\beann}{\begin{eqnarray*}}
\newcommand{\eeann}{\end{eqnarray*}}
\newcommand{\beasn}{\begin{sneqnarray}}
\newcommand{\eeasn}{\end{sneqnarray}}

\newcommand{\al}{\alpha}

\newcommand{\AmS}{{\protect\the\textfont2
  A\kern-.1667em\lower.5ex\hbox{M}\kern-.125emS}}

\title{Effective Field Theory for Ultrasoft Momenta in NRQCD and NRQED} 

\author{A. Pineda and J. Soto \address{
          Departament d'Estructura i Constituents
          and Institut de F\'\i sica d'Altes Energies.\\
          Universitat de Barcelona, Diagonal, 647
          E-08028 Barcelona, Catalonia, Spain.}%
        \thanks{A.P. acknowledges a fellowship from Generalitat of Catalunya.
 Financial support from CICYT, contract AEN95-0590 and financial
support from CIRIT, contract GRQ93-1047 is also acknowledged.}}
       
\begin{document}

\begin{abstract}
We propose an effective field theory for heavy quark-antiquark bound states 
once the soft gluons have been integrated out. We also give new results for 
the matching between QCD (QED) and NRQCD (NRQED).
\end{abstract}

\maketitle

\section{INTRODUCTION}

Two particle non-relativistic bound states in QCD (QED) have, at least, 
three well 
separated scales: the mass $m$ (hard scale), the typical relative momentum $|\vec p|$
(soft scale) and the typical bound state energy $E$ (ultrasoft scale). This 
allows to introduce a hierarchy of effective field theories when sequentially 
integrating out each of these scales. After integrating out the hard scale 
$m$ QCD (QED) becomes a non-relativistic theory, the so called NRQCD (NRQED) 
\cite{Lepage}. 
This effective theory is local in space and time and it is still naturally 
written in terms of quark fields.
Integrating out the soft scale leads to a so far elusive effective field 
theory which we call potential NRQCD (pNRQCD). We claim that this effective 
theory is 
local in time but it is non-local in space and it is naturally written in 
terms of wave function fields. 

We present some original results for the above mentioned effective field theories. 
In section 2
we give the matching coefficients for the four-quark operators at one loop for NRQED
and also for NRQCD in the case of different quark masses. 
In section 3 we put forward our proposal for pNRQCD and give some results for 
pNRQED.

\section{FROM QCD TO NRQCD}

The matching for NRQCD (and HQET) has been known at tree level since
long. This can be obtained doing the matching at tree level of S-matrix
elements or just performing a F-W transformation of the QCD lagrangian.
Some results at one loop have also been known for HQET.
Nevertheless, attempts to perform the matching
beyond tree level using dimensional regularisation (DR)
 in NRQCD have not appeared until recently. The problem
was that in NRQCD unlike in HQET the kinetic term was
incorporated in the propagator. In DR the high modes are not explicitly suppressed
by a cut-off $\mu$ ($\mu << m$) and give non-vanishing contributions which break the
power counting rules.

Lately several people have addressed this problem \cite{res,otros,Manohar} 
and recently the situation has been clarified \cite{Manohar}. There, it
is claimed that the matching should be performed just like in HQET. Let
us
make some comments in favor of this approach. The key point is that when
doing the matching it is not so important to know the power counting in the effective
theory as
to know that the scales of the effective theories are much lower than
the
mass. The power counting will tell us the relative importance between
different operators but this would not change the value of the matching
coefficients. That is, we only need 

\be
m >> |\vec p|,\, E,\, \Lambda_{QCD}
\ee
whatever the relation between $|\vec p|$, $E$ and $\Lambda_{QCD}$ is.

In ref. \cite{Manohar} DR was used
for both ultraviolet (UV) and infrared (IR) divergences in the full and
the
effective theory. In fact, it is not so important to know the way the UV
divergences of the full theory are regulated since the comparison is
done between S-matrix elements which are UV finite. Nevertheless, it is
essential to regulate in the same way the IR divergences of both
theories in order to cancel. This will happen since both theories have
by construction
the same IR behavior. It is also very important, from a practical
point of view, to regulate the UV
divergences of the effective theory using DR.
In this way,
in ref. \cite{Manohar} 
the calculation in the effective theory becomes
trivial, since there is no dimensionful parameter in the integrand, and
 the matching coefficients for the bilinear terms in fermions 
are calculated at one loop. 
 Nevertheless, the four-quark operators were not taken into
account. The way to deal with these operators is not obvious since we are 
faced 
with the computation of S-matrix elements of
four-quarks in QCD and HQET. It is in these S-matrix elements, which are 
never calculated in traditional applications of HQET, where the 
distinct 
 IR behavior
of two heavy quark bound states becomes apparent. Power-like IR divergences appear 
in loops where a quark and an antiquark in HQET interact through a potential. 
We call these divergences Coulomb pole. They are naturally regulated once 
the kinetic term is introduced.

This IR behavior should appear in both the full and the effective theory. 
However, 
it is important to bear in mind 
that the matching coefficients are independent
of the IR behavior. Therefore, we do not need to regulate the IR singularity 
with the introduction of the kinetic term and hence we can take advantage of 
a more convenient regularisation. For instance, a regularisation such that 
we could avoid the effort of computing this pole in
both theories (which can be very painful).

The procedure consist of computing the matrix elements on-shell and at 
threshold 
($|\vec p| =0$). In this way there is no scale in the effective theory, and hence 
the diagrams in the effective theory are just trivially zero. 
Therefore, only the diagrams in QCD need to be computed 
in this 
peculiar
kinematical situation. This precise kinematical 
situation produces IR divergences which get canceled by those of the 
effective theory. The computation is done in the $\overline{MS}$ scheme.

We stress that the Coulomb pole does not appear at all
doing the matching in this way. Let us explain what happens. In order to 
define some integrals we have to
move to dimensions high enough for the IR Coulomb
singularity to be regulated. When coming back to four dimensions we can trace back the
IR
Coulomb singularity as poles in dimensions different from four.
 The point is that we have not introduce
the relative momentum and hence DR has no
way to reproduce the Coulomb pole or any non-local behavior in the relative 
momentum. This fact has already been observed 
in ref.
\cite{nos2}, where it was noticed that in HQET with a quark and antiquark
moving exactly at the same velocity no imaginary anomalous dimensions
occur using DR.

The important thing doing the matching is to take into account all the
non-analytical behavior which cannot be obtained in the effective theory. We
are taking into account all the non-analytical behavior due to the masses.
The remaining non-analytical behavior is encoded in the effective theory.

Let us now give the results for the four-quark effective lagrangian 
(non-equal mass case)
\bea
\label{lag1}
\nonumber
&&\delta {\cal L}_{NRQCD} =
  {d_{ss} \over m_1 m_2} \psi_1^{\dag} \psi_1 \chi_2^{\dag} \chi_2
\\
&&
\nonumber
+
  {d_{sv} \over m_1 m_2} \psi_1^{\dag} {\vec \sigma} \psi_1
                         \chi_2^{\dag} {\vec \sigma} \chi_2
+
  {d_{vs} \over m_1 m_2} \psi_1^{\dag} {\rm T}^a \psi_1
                         \chi_2^{\dag} {\rm T}^a \chi_2
\\
&&
+
  {d_{vv} \over m_1 m_2} \psi_1^{\dag} {\rm T}^a {\vec \sigma} \psi_1
                         \chi_2^{\dag} {\rm T}^a {\vec \sigma} \chi_2
\,,
\eea

\bea
&&
\nonumber
d_{ss}=
  - {N^2_c-1 \over 4N^2_c} {\al_s^2 \over m_1^2-m^2_2}
\Biggl\{m_1^2\left(  \ln{m^2_2 \over  \nu^2}
                   + {1 \over 3} \right)
\\
&&
       -
       m^2_2\left(  \ln{m^2_1 \over  \nu^2}
                   + {1 \over 3} \right)
\Biggr\}
\eea
\be
d_{sv}=
   {N^2_c-1 \over 4N^2_c} {\al_s^2 \over m_1^2-m^2_2}
m_1 m_2\ln{m^2_1 \over m^2_2}
\ee
\bea
\label{dvs}
\nonumber
&&d_{vs}=
- {2 C_f \al^2_s \over m_1^2-m^2_2}
  \Biggl\{m_1^2\left( \ln{m^2_2 \over  \nu^2}
                   + {1 \over 3} \right)
\\
&&
       -
       m^2_2\left(  \ln{m^2_1 \over  \nu^2}
                   + {1 \over 3} \right)
  \Biggr\}
\\
&&
\nonumber
+ { C_A \al^2_s \over 4 (m_1^2-m^2_2)}
 \Biggl[
  3\Biggl\{m_1^2\left( \ln{m^2_2 \over \nu^2}
                   + {1 \over 3} \right)
\\
&&
\nonumber
       -
       m^2_2\left(  \ln{m^2_1 \over  \nu^2}
                   + {1 \over 3} \right)
  \Biggr\}
\\
&&
\nonumber
  +
   { 1 \over m_1m_2}
   \Biggl\{m_1^4\left( \ln{m^2_2 \over \nu^2}
                   + {10 \over 3} \right)
\\
&&
\nonumber
       -
       m^4_2\left(  \ln{m^2_1 \over  \nu^2}
                   + {10 \over 3} \right)
  \Biggr\}
 \Biggr]
\eea
\bea
\label{dvv}
&&d_{vv}=
 {2 C_f \al^2_s \over m_1^2-m^2_2}
        m_1 m_2\ln{m^2_1 \over m^2_2}
\\
&&
\nonumber
+
 { C_A \al^2_s \over 4 (m_1^2-m^2_2)}
    \Biggl[
  \Biggl\{m_1^2\left( \ln{m^2_2 \over  \nu^2}
                   + 3 \right)
\\
&&
\nonumber
       -
     m^2_2\left(  \ln{m^2_1 \over \nu^2}
                   + 3 \right)
  \Biggr\}
    -
  3 m_1 m_2\ln{m^2_1 \over m^2_2}
    \Biggr]
\,,
\eea

where
$$
C_f = {N^2_c-1 \over 2N_c} \quad \quad {\rm and} \quad \quad C_A=N_c \,.
$$

The QED coefficients are easily obtained from these results. We just have to 
omit $d_{vs}$ and $d_{vv}$ and replace ${N^2_c-1 \over 4N^2_c}$ by $1$.

In the equal mass case annihilation processes are allowed and they should be 
taken into account. For QED, joining all the contributions we get

\be
\label{dssqed}
d_{ss}=
{3 \pi \al \over 2}
\Biggl\{ 1
  - {2 \al \over 3 \pi}
\left(  \ln{m^2 \over  \nu^2}
                   + {23 \over 3} - \ln2 + i {\pi \over 2} \right)
\Biggr\}
\ee
\be
\label{dsvqed}
d_{sv}=
-{ \pi \al \over 2}
\Biggl\{ 1
  - {2 \al \over  \pi}
\left(  {22 \over 9} + \ln2 - i {\pi \over 2} \right)
\Biggr\} \, .
\ee
These results are compatible with those found by Labelle et al. \cite{Labelle} 
except for a finite piece in (\ref{dssqed}).

A more detailed explanation of the procedure and the full results at one 
loop for the equal and non-equal mass case in QCD will be given in ref. 
\cite{nos9}.

\section{FROM NRQCD TO pNRQCD}

In the last section we have integrated out the hard gluons. Here, 
we integrate out soft gluons, with energies of the order of the relative 
momentum.

Two point functions are insensitive to the relative momentum and hence the bilinear
terms in fermions in the NRQCD lagrangian and in the pNRQCD at quark level lagrangian 
will read
 exactly the same. However one has to keep in mind that in the latter only
gluons with ultrasoft momenta are kept. On the contrary, four point functions 
do know about relative
momentum and generate non-trivial terms in the pNRQCD lagrangian. These terms are
nothing but 
 the potential piece of the new lagrangian. 
Due to the massless nature of the gluons the coefficients are going to be 
non-local in the relative space coordinate although local in time. The 
important point to be realized is that the appearance of a
potential can be understood as the effect of integrating out soft gluons. Hence the
potential can be calculated by matching NRQCD to pNRQCD.

We have carried out the matching to a given order in $1/m$ and 
$\al$ using 
HQET propagators and the Coulomb gauge. This produces a very strong simplification since the kinetic 
term can be treated perturbatively when computing the potential. Now, it is 
very easy to know how far we must go in the computation of the potential if 
we want to compute, for instance, the energy up to order $m\al^n$. In this case, we must 
compute the matching up to order $({1 \over m})^s\al^r$, with $s$, $r$ such 
that $s+r \leq n-1$. The lowest order just gives the standard Schr\"odinger 
equation. 

In addition we use DR for
both UV and IR divergences, and hence any loop in pNRQCD gives zero because there is no
scale. The point is that on the one hand there is no way of 
reproducing the Coulomb pole since the computation is done with HQET 
propagators in both theories. On the other hand any loop with 
ultrasoft gluons in pNRQCD is zero since these are only sensible to the incoming energy 
and total momentum of the S-matrix element, which we set to zero (we remark that 
on-shell quarks in HQET have zero energy).
Therefore, the calculation from 
NRQCD leads directly to the potential. 
Further terms in the pNRQCD lagrangian are obtained when matching four point 
functions with an arbitrary number of ultrasoft gluon legs.

In order to make explicit the distinction between soft and ultrasoft gluons it is most
convenient to project NRQCD or pNRQCD to the two particle sector and promote the wave
function $\psi (\vec x_1 ,\vec x_2 , t) $ to a field ($\psi (\vec x_1 ,\vec x_2 , t) $
 is a $3\times 3$ matrix
in color space and a $2\times 2$ matrix in spin space). Then the relative coordinate
$\vec x =\vec x_1 -\vec x_2$, whose typical size is the inverse of the soft scale, is
explicit and can be consider a small scale compared with the typical wavelength of the 
ultrasoft gluon which is also of the order of the inverse of the energy $E$. 
Gluon fields appear in this 
formalism at
the points $\vec x_1$ and $\vec x_2$. Ultrasoft gluons are those which $B_{\mu}
(\vec x_{i},\, t)$
can be expanded about the center of mass coordinate in a power series of derivatives
and relative coordinates, the so-called multipole expansion (refs. \cite{otros,mp}
also deal with ultrasoft momenta through the multipole expansion). Then the most natural
 way of writing the pNRQCD lagrangian
is as a functional of $\psi (\vec x ,\vec X , t) $ and $B_{\mu}
(\vec X , \, t)$, which is local in $\vec\nabla_{\vec x}$, 
$\vec\nabla_{\vec X}$ but non-local in $\vec x$.  For definiteness we will 
give $L_{pNRQCD}$ at the leading order in the potential.
\bea
\label{lpnrqcd} 
&&
{\cal L}_{pNRQCD} =
\int d^3{\vec x} d^3{\vec X} dt tr \biggl(\psi^{\dagger} (\vec x_1 ,\vec x_2 , t)
\\
&&
\nonumber
\Bigl\{
iD_0  +{\vec D_{\vec x_1 }^2\over 2m}+
{\vec D_{\vec x_2 }^2\over 2m}
\Bigr\}\psi (\vec x_1 ,\vec x_2 , t)\biggr)
\\
&&
\nonumber
+{\al \over \vert  x_1 - x_2 \vert }tr \biggl(T^{a}
\psi (\vec x_1 ,\vec x_2 , t)T^{a}\psi^{\dagger} (\vec x_1 ,\vec x_2 , t)\biggr)
\,.
\eea

Recall that the gluon fields in the covariant derivatives are ultrasoft and 
hence they must be multipole expanded which spoils the explicit gauge 
invariance.
However, this can be recovered by writing $\psi (\vec x ,\vec X , t) 
$ in terms of the scalar
 $S (\vec x ,\vec X , t) $ and octet $O (\vec x ,\vec X , t) $ wave function fields,
whose local gauge transformations depend on the center of mass coordinate only. Namely,
\bea
&&
\nonumber
\psi (\vec x_1 ,\vec x_2 , t)= P\bigl[e^{ig\int_{\vec x_2}^{\vec x_1} \vec B 
                   d\vec x} \bigr]S({\vec x}, {\vec X}, t)
\\
&&
\nonumber
  +P\bigl[e^{ig\int_{\vec X}^{\vec x_1} \vec B d\vec x}
\bigr]O (\vec x ,\vec X , t)P\bigl[e^{ig\int^{\vec X}_{\vec x_2} \vec B d\vec x}
\bigr]
\eea
$$\psi (\vec x_1 ,\vec x_2 , t)\rightarrow g(\vec x_1 ,t)\psi (\vec x_1 ,\vec x_2 , t)
g^{-1}(\vec x_2 ,t ) $$
$$S (\vec x ,\vec X , t)\rightarrow S (\vec x ,\vec X , t) $$
$$O (\vec x ,\vec X , t)\rightarrow g(\vec X ,t)O (\vec x ,\vec X , t)g^{-1}(\vec X
,t) 
 $$
In this way (\ref{lpnrqcd}) reads
\bea
&&{\cal L}_{pNRQCD} =
\int d^3{\vec x} d^3{\vec X} dt tr \Biggl\{
\\
&&
\nonumber
 S^{\dagger}
              \Bigl\{
i\partial_0 - { {\vec p}^2 \over m} + {C_{f} \al \over |{\vec x}|} 
\Bigr\} S
\\
&&
\nonumber
+ O^{\dagger}
                \Bigl\{
iD_0 - { {\vec p}^2 \over m} - {1 \over 2N_c} {\al \over |{\vec x}|}
\Bigr\} O 
\\
&&
\nonumber
+g\vec x  O \vec E (\vec X , t)
S^{\dagger}
+g\vec x  O^{\dagger} \vec E (\vec X , t)
S
\\
&&
\nonumber
+{g\over 2}\vec x  O O^{\dagger} \vec E (\vec X , t)
+{g\over 2}\vec x  O^{\dagger} O \vec E (\vec X , t) \Biggr\}
\,.
\eea
This lagrangian suffices to obtain 
the leading non-perturbative
contributions to the two heavy quark bound states when $E >> \Lambda_{QCD}$ 
\cite{VL} (see also \cite{yndnos}).

If we leave aside non-perturbative effects ($\sim \Lambda_{QCD}$), each term in 
the lagrangian above has a well defined size, unlike in the NRQCD lagrangian. Relative
coordinates $\vec x $ and its associated momentum $\vec\nabla_{\vec x}$ must be
 counted as
soft scales ($(\vec x )^{-1}\sim\vec\nabla_{\vec x}\sim m\alpha $). Gluon fields
  $B_{\mu}
(\vec X ,t)$ and derivatives with respect to the center of mass coordinate 
 $\vec\nabla_{\vec X}$ must be counted as ultrasoft scales ($\sim m\alpha^2 $).
 Then if we
wish to calculate a given observable to a given order in $\alpha$ we know immediately 
which terms are to be kept in the pNRQCD lagrangian. As an example, let us present  
 $L_{pNRQED}$ from which one can obtain next to next to
leading corrections to the energy ($\sim m\alpha^5 $).
\bea
\label{lpnrqed}
\nonumber
&&{\cal L}_{pNRQED} =
\int d^3{\vec x} d^3{\vec X} dt S^{\dagger}({\vec x}, {\vec X}, t)
\\
&&
\nonumber
                \Biggl\{
i\partial_0 - { {\vec p}^2 \over m} + { \al \over |{\vec x}|} 
+ { {\vec p}^4 \over 4m^3}
\\
&&
\nonumber
- { \delta^{(3)}({\vec x}) \over m^2} 
       \left( \pi \al \left(c_D -2c_F^2 \right) +d_{ss}+3d_{sv} \right)
\\
&&
\nonumber
+ { \al \over 2 m^2} { 1 \over {\vec x}}
       \left( {\vec p}^2 + { 1 \over {\vec x}^2} {\vec x} 
                 ({\vec x} \cdot {\vec p}){\vec p} \right)
\\
&&
\nonumber
- { \delta^{(3)}({\vec x}) \over m^2} {\vec S}^2
       \left( \pi \al { 4 \over 3}c_F^2 -2 d_{sv} \right)
\\
&&
\nonumber
- { \al \over 4 m^2} { 1 \over |{\vec x}|^3} {\vec L} \cdot {\vec S}
       \left( 2c_S+4c_F \right)
\\
&&
\nonumber
- { \al c_F^2 \over 4 m^2} { 1 \over |{\vec x}|^3} 
            S_{12} ({\vec x})
- \delta V ({\vec x}) 
+ e {\vec x} \cdot {\vec E} ({\vec X},t)
\Biggr\}
\\
&&
S ({\vec x}, {\vec X}, t)
\,.
\eea
The coefficients $c_{i}$ are given in \cite{Manohar} and $d_{ss}$, $d_{sv}$ 
in (\ref{dssqed}) and (\ref{dsvqed}).
 $c_{i}=1+O(\alpha )$ and $d_{ij}=O(\alpha )+O(\alpha^2 ) $ are obtained from the
one loop matching between QED and NRQED. The last term corresponds 
to the  
ultrasoft photons which contribute at this order. The potential terms are 
obtained upon
matching NRQED to pNRQED up to one loop. $\delta V$ encodes the computation 
of the potential at one loop. It reads
\bea
\nonumber
&&\delta V ({\vec x}) = - {\al^2 \over m^2}
   \left(\delta^{(3)}({\vec x}) \ln\nu^2+ {1 \over 2\pi} {\rm reg}
 {1 \over |{\vec x}|^3} \right)
\\
&&
\nonumber
   - { 4 \al^2 \over  3 m^2}
   \left(\delta^{(3)}({\vec x}) \ln\nu^2+ {1 \over 2\pi} {\rm reg}
 {1 \over |{\vec x}|^3} \right)
\\
&&
   -  {\al^2 \over m^2} \delta^{(3)}({\vec x}) C
\eea
with unknown $C$. The first $\ln\nu$ (which was produced by an UV 
divergence in the potential)  
gets canceled by the $\ln\nu$ in $d_{ss}$. The second $\ln\nu$ (with IR 
origin) cancels with a piece of the UV divergent contribution coming from 
the ultrasoft photons. The remaining contribution from the ultrasoft photons 
cancels the contribution coming from $c_D$. The net result is the total 
energy to be scale independent as it must be.
 
With (\ref{lpnrqed}) the binding energy at order $m\alpha^5$ for
arbitrary $n,l$ states can be calculated. We find agreement with 
\cite{Labelle} for the hyperfine splittings. The $m \al^5\ln\al$ correction 
has also been 
calculated finding agreement with known results \cite{Gupta}. 
We have
also used these techniques with DR to reproduce the Lamb shift in the 
simpler case of a hydrogen-like atom \cite{nos8}.

We remark that in this section we are assuming $|\vec p | >> E$ which was 
not needed in the 
previous section. As far as we are making the matching to some order in $\al$ 
(we are computing the potential perturbatively) we also assume $|\vec p | 
>> \Lambda_{QCD}$. Notice that the relative size between $E$ and 
$\Lambda_{QCD}$ is left arbitrary.  We can further distinguish between
 two
situations: (i) $E >> \Lambda_{QCD}$ and (ii) $\Lambda_{QCD} 
{\ \lower-1.2pt\vbox{\hbox{\rlap{$>$}\lower5pt\vbox{\hbox{$\sim$}}}}\ } E$. 
For 
the situation (i), as we mention before, we can calculate perturbatively 
from pNRQCD and 
parametrise the non-perturbative contributions
by means of local condensates. For the situation (ii) the calculations in 
pNRQCD
cannot be carried out perturbatively anymore. 

If $\Lambda_{QCD} 
{\ \lower-1.2pt\vbox{\hbox{\rlap{$>$}\lower5pt\vbox{\hbox{$\sim$}}}}\ } 
|{\vec p}|$ the matching to
pNRQCD cannot be carried out perturbatively. Even in this situation NRQCD is
extremely useful to parametrise
non-perturbative contributions in many processes \cite{Lepage}.



\end{document}